\documentclass[aps,prev,twocolumn,preprintnumbers,floatfix,nofootinbib]{revtex4}
\pdfoutput=1 

\usepackage{amsmath}
\usepackage{amsfonts}
\usepackage{amssymb}
\usepackage{xcolor}
\usepackage{graphicx}
\usepackage{bm}
\usepackage{times}
\usepackage{hyperref}

\begin{document}

\preprint{IIPDM-2019}


\title{\boldmath Explaining the AMS positron excess via Right-handed Neutrinos}



\author{Farinaldo S. Queiroz}
\email{farinaldo.queiroz@iip.ufrn.br}
\affiliation{International Institute of Physics, Universidade Federal do Rio Grande do Norte,
Campus Universitario, Lagoa Nova, Natal-RN 59078-970, Brazil}

\author{Clarissa Siqueira}
\email{csiqueira@iip.ufrn.br}
\affiliation{International Institute of Physics, Universidade Federal do Rio Grande do Norte,
Campus Universitario, Lagoa Nova, Natal-RN 59078-970, Brazil}

\begin{abstract}
We have witnessed in the past decade the observation of a puzzling cosmic-ray excess at energies larger than $10$~GeV. The AMS-02 data published this year has new ingredients such as the bump around $300$~GeV followed by a drop at $800$~GeV, as well as smaller error bars. Adopting the background used by the AMS-02 collaboration in their analysis, one can conclude that previous explanations to the new AMS-02 such as one component annihilating and decaying dark matter as well as pulsars seem to fail at reproducing the data. Here, we show that in the right-handed neutrino portal might reside the answer. We discuss a decaying two-component dark matter scenario where the two-body decay products are right-handed neutrinos that have their decay pattern governed by the type I seesaw mechanism.  This setup provides a very good fit to data, for example, for a conservative approach including just statistical uncertainties leads to $\chi^2/d.o.f \sim 2.3$ for $m_{DM_1}=2150$~GeV with $\tau_{1}=3.78 \times 10^{26}$~s and $m_{DM_2}=300$ with $\tau_{2}=5.0 \times 10^{27}$~s for $M_N=10$~GeV, and, in an optimistic case, including systematic uncertainties, we find $\chi^2/d.o.f \sim 1.12$, for $M_N = 10$~GeV, with $m_{DM_1}=2200$~GeV with $\tau_{1}=3.8 \times 10^{26}$~s and $m_{DM_2}=323$~GeV with $\tau_{2}=1.68 \times 10^{27}$~s.
\end{abstract}

\maketitle
\flushbottom

\section{Introduction}
\label{sec:intro}

The observation of cosmic-rays have boosted our understanding of astrophysical phenomena that undergo diffusion and energy loss processes in the intergalactic medium. Historically, in 2008 the Payload for Antimatter Matter Exploration and Light-nuclei Astrophysics (PAMELA) surprisingly announced the first evidence of a rise in the cosmic-ray positron fraction at GeV energies with high statistics \cite{Adriani:2008zr}.  Fermi-LAT confirmed this cosmic-ray anomaly much later in 2011. Taking advantage of the absent onboard magnet, they could distinguish electrons from positrons by exploiting the Earth's shadow, which is offset in opposite directions for opposite charges due to the Earth's magnetic field. With this technique they were able to indeed observe a positron fraction rise for energies between $20$ and $200$~GeV \cite{FermiLAT:2011ab}. With much better statistics, the AMS mission measured the positron fraction up to $350$~GeV \cite{Aguilar:2013qda}, and reported a flat positron fraction for energies above $150$~GeV. 

That has triggered a number of works which were able to explain the AMS excess of events. Some attempts focused on annihilating dark matter \cite{Cholis:2013psa,Ibarra:2013zia}, but the annihilation cross section needed to fit the excess was too large to be in agreement with gamma-ray observations in the direction of the galactic center and dwarf spheroidal galaxies \cite{Hooper:2012sr,Ackermann:2015zua}, and cosmic microwave background data \cite{Galli:2011rz,Weniger:2013hja}.  Interpretations in terms of decaying dark matter were also put forth, where a lifetime of the order of $10^{27}$~s for $\mu\bar{\mu}$ final states could provide a reasonable fit to data \cite{Nardi:2008ix,Arvanitaki:2008hq,Dienes:2013xff,Geng:2013nda,Belotsky:2014haa}. Alternatively, nearby astrophysical objects presented themselves as good candidates \cite{Profumo:2008ms, Hooper:2008kg,Grasso:2009ma}. That was the whole story until the new AMS data and HAWC observations came into light.

The new AMS data has new ingredients \cite{Aguilar:2019owu}: (i)
features much smaller error bars at low energies and a rise at $\sim 10$~GeV; (ii) the previously observed flat spectrum for energies larger than $150$~GeV now exhibits a bump-like feature with a peak around $300$~GeV; (ii) a sharp drop for energies above $400$~GeV is visible. These new ingredients significantly harden the shape of the spectrum making a dark matter interpretation difficult, especially adopting the single component scenario. Moreover, the High-Altitude Water Cherenkov Observatory (HAWC) observed the presence of energetic electrons and positrons from nearby pulsars and from that the diffusion parameters were inferred. The diffusion parameters derived are inconsistent with the one observed by AMS-02 though, thus ruling out such pulsars as the origin of the AMS excess \cite{Abeysekara:2017old}. In conclusion, the new AMS data begs for a new interpretation \cite{Farzan:2019qdm}.

In this work, we attempt to explain the positron excess in terms of two-component dark matter comprised of two scalars. Such scalars decay into two right-handed neutrinos that decay into Standard Model particles according to the type-I seesaw mechanism \cite{Campos:2017odj,Batell:2017rol}. This scenario appears in Majoron-inspired models, for instance \cite{Gelmini:1982rr,Gelmini:1984pe,Santamaria:1986kg,Choi:1991aa,Berezinsky:1993fm,Chang:2014lxa,Queiroz:2014yna,Boucenna:2014uma,Ma:2017xxj,Garcia-Cely:2017oco,Brune:2018sab}. We emphasize that in the canonical Majoron model, the decay into right-handed neutrinos is not dominant.  Decays into left-handed neutrinos are instead more relevant, and they lead to an interesting phenomenology explored elsewhere \cite{Garcia-Cely:2017oco}. In this work, we are investigating the possibility of fitting the AMS-02 data with a two-component dark matter setup where each component decays into two right-handed neutrinos. We are not interested in a explicit theoretical realization of this scenario but we do emphasize that having a two-component decay dark matter model requires going beyond the vanilla Majoron models and other type I seesaw model incarnations. Our idea is simply to assess whether one could get a reasonable fit to the AMS-02 data if such decays are dominant, without having an specific model at hand.

That said, we perform a chi-squared analysis choosing different masses for the right-handed neutrino ($10$~GeV, $50$~GeV and $80$~GeV) and leaving the DM mass and the decay rate as free parameters to get the best fit to the data. In addition, we choose two different set of propagation parameters which are known as {\it medium} (MED) and {\it maximum} (MAX) diffusion models, using the Navarro-Frenk-White (NFW) profile.

Moreover, we carry out all this procedure including only statistical errors, and statistical plus systematic errors to really assess the impact of the systematic effects on our conclusions. Including only the statistical uncertainties we find the best-fit of $\chi^2/d.o.f \sim 2.3$ for $m_{DM_1}=300$ with $\tau_{ 1}=1.67 \times 10^{27}$~s and $m_{DM_2}=2000$~GeV with $\tau_{DM_2}=4 \times 10^{26}$~s for $M_N=10$~GeV, and, for the optimistic case, including systematic uncertainties, we get $\tau_{ 1}=1.68 \times 10^{27}$~s and $\tau_{DM_2}=3.8 \times 10^{26}$~s, for $m_{DM_1}=323$~GeV, and $m_{DM_2}=2200$~GeV respectively with $M_N = 10$~GeV, yielding $\chi^2/d.o.f \sim 1.12$.

Lastly, we put our results into perspective with gamma-rays observations \cite{Ando:2015qda,Massari:2015xea}. We start our reasoning discussing below how we obtain the positron flux.

\section{Positron flux}
\label{sec:dmann}
The positron flux reported by AMS seems to be compatible with a background, which is given by a diffuse flux at low energies and a new source at high energies. So, the collaboration interpreted the whole signal as a background plus a new source term as follows,
\begin{equation}
    \Phi^{e^+}_{tot} = \Phi_{diffuse}^{e^+} + \Phi_{source}^{e^+}.
\end{equation}
In this work, we choose decaying dark matter particles to be responsible for this new source flux, $\Phi_{source}$, described above. For this purpose, it is necessary to compute the decaying DM positron flux, which is given by,
\begin{widetext}
\begin{equation}
    \Phi^{e^+}_{DM} (E) = \frac{1}{4\pi b(E)} \frac{\rho_\odot}{m_{DM}} \Gamma \times \int_E^{m_{DM}/2}dE_s \sum_{f} BR_f \frac{dN^{e^+}_f}{dE}(E_s)\mathcal{I}(E,E_s) 
    \label{posflux}
\end{equation}
\end{widetext}
where $E$ is the positron energy after propagation and $E_s$ is the positron energy at production, $\rho_\odot = 0.4$~GeV/cm$^3$ is the DM density in the location of the Sun, $m_{DM}$ is the DM mass, $\Gamma$ is the decay rate of DM particle, $BR_f$ is the branching ratio for a given final state $f$ and $\frac{dN^{e^+}_f}{dE}(E_s)$ is the number of positrons per energy produced after decay before the propagation. The parameter $b(E)$ is the called energy loss function, which takes into account the possible energy losses via synchrotron radiation and inverse Compton scattering.

For the purpose of being conservative, we choose the same diffuse flux as reported by the collaboration which includes contributions from the interaction between galactic cosmic rays with the intergalactic medium, 
\begin{equation}
    \Phi_{diffuse}^{e^+}(E)= c_d\frac{E^2}{\hat{E}^2} \left( \frac{\hat{E}}{E_1} \right)^{\gamma_d}
    \label{backflux}
\end{equation}
where the values for the parameters reported by the collaboration were: $E_1=7$~GeV, $\hat{E} (E)= E + \varphi_{e^+}$, with $\varphi_{e^+}=1.10 \pm 0.03$~GeV, $c_d=(6.51 \pm 0.14 )\times 10^{-2} ({\rm m^2\,sr\,s\,GeV})^{-1}$,  $\gamma_d=-4.07 \pm 0.06$, where we use the central values for the parameters $E_1$ and $\varphi_{e^+}$, while the values for $c_d$ and $\gamma_d$ were chosen within $3\sigma$ contour in order to provide the best-fit to the data.

Furthermore, the halo function $\mathcal{I}(E,E_s)$, computed using the numerical package PPPC4DMID, appears as a solution to the diffusion equation, and it is dependent on the loss energy function ($b(E)$), on the DM profile (here we choose the NFW), on the diffusion parameters $\mathcal{K}_0 = 0.0112$~kpc$^2/$~Myr and $\delta = 0.70$ for {\it medium} (MED) and $\mathcal{K}_0 = 0.0765$~kpc$^2/$~Myr and $\delta = 0.46$ for the {\it maximum} (MAX) propagation models.

In the next, we will compute the fluxes for the model considered here.

\section{Results}
\label{sec:results}

The scenario involves two DM particles decaying into two right handed neutrinos (RHN) pairs. We assume each DM candidate composing $50\%$ of the DM abundance of the Universe, and of course these values can be easily changed by rescaling the decay rate accordingly. These RHN couples to standard model particles via Higgs and gauge bosons, leading to the following RHN decay pattern $N_R \rightarrow W^{+/-} + l^{-/+}$, $N_R \rightarrow Z + \nu_l$, and $N_R \rightarrow H + \nu_l$, in principle, $l$ can be the three leptonic flavors, but in our case,
for simplicity, we choose the $l=e$. In addition, we impose three different values for the RHN mass, $M_N=10$~GeV, $M_N=50$~GeV, and $M_N=80$~GeV.

\begin{figure}[ht!]
    \centering
    \includegraphics[width=0.45\textwidth]{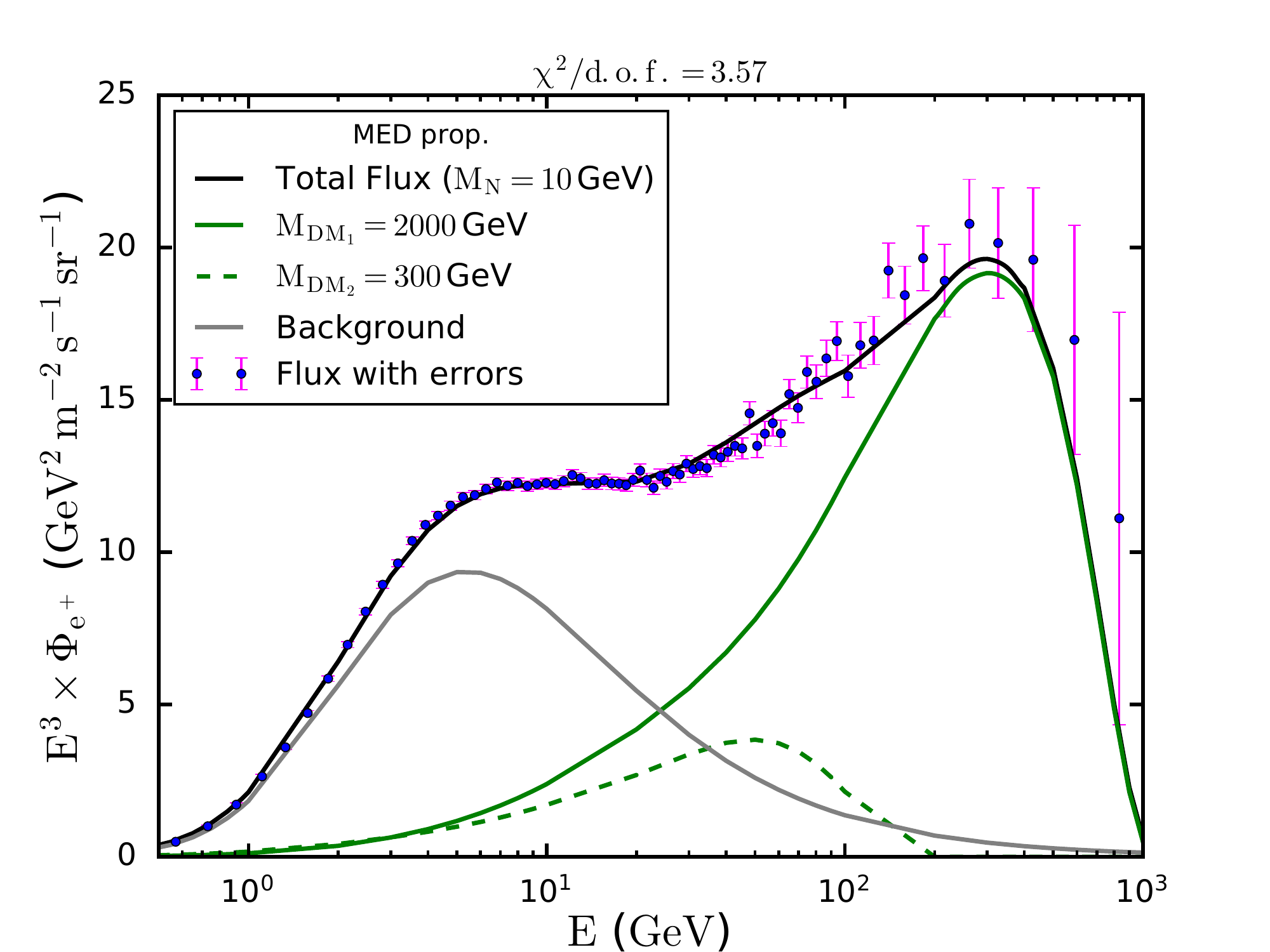}
    \includegraphics[width=0.45\textwidth]{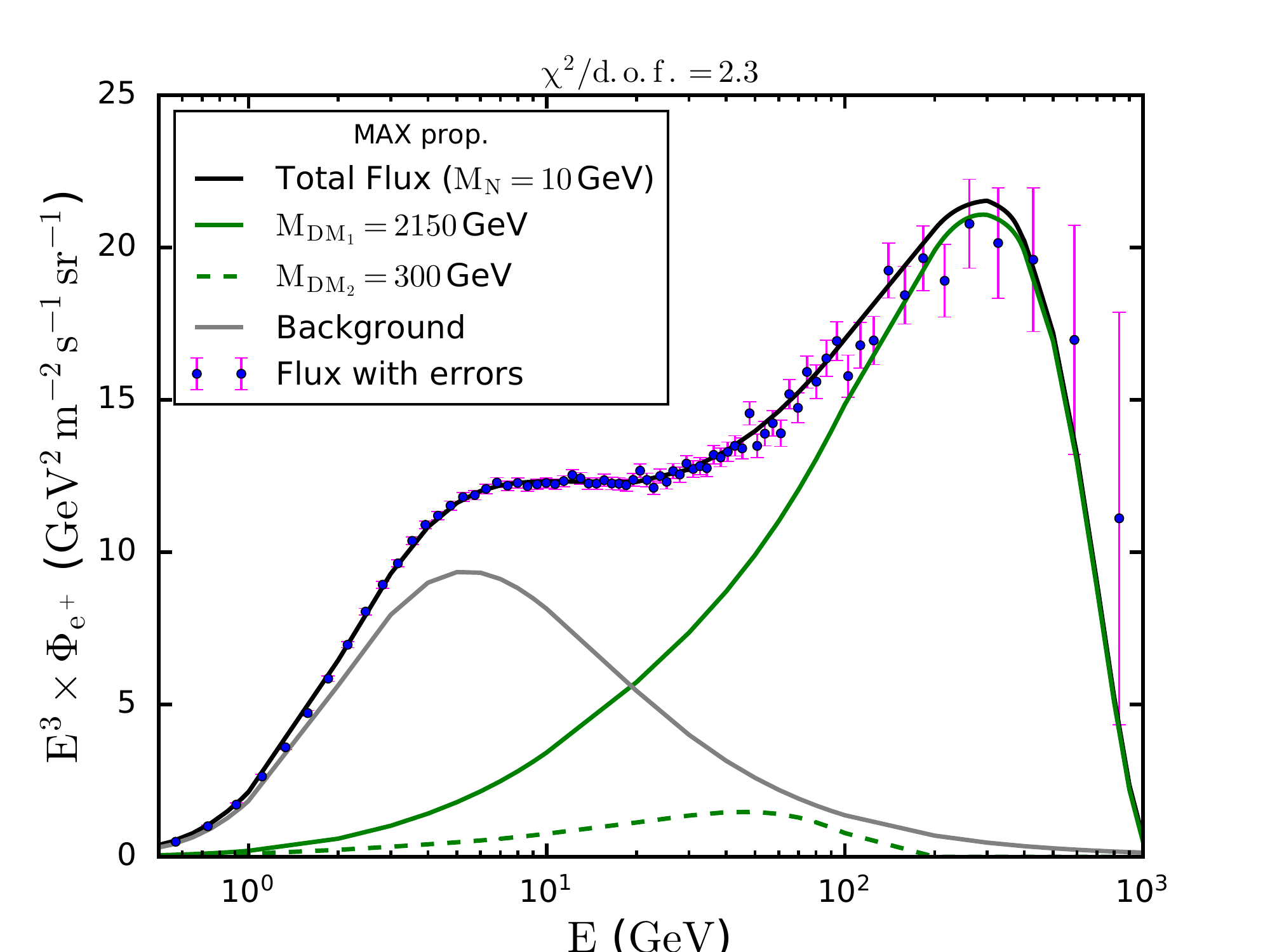}
    \caption{Positron flux \textit{versus} energy, summing over different contributions (black line): Dark Matter candidate 1 (continuous green line), Dark Matter candidate 2 (dashed green line), Diffuse Background (gray line). In this case we choose the right-handed neutrino mass equal to $10$~GeV for two different propagation models, MED (top) and MAX (bottom), with $\chi^2/d.o.f.=3.57$ and $\chi^2/d.o.f.=2.3$, respectively. }
    \label{fig:flux10MED}
\end{figure}

Fixing the RHN masses, we compute the positron flux in Eq.~(\ref{posflux}) using the PPPC4DMID code which computes the halo function $\mathcal{I}(E,E_s)$, and the Pythia 8 package to obtain the positron spectrum for each right-handed neutrino mass. Then we left as free parameters the DM masses $M_{DM_1}$ and $M_{DM_2}$ and the decay rates in order to fit the data reported by the AMS collaboration, namely, for each RHN mass we found a combination of DM mass \textit{versus} decay rate which provides the best-fit for the data. 

\begin{figure}[ht!]
    \centering
    \includegraphics[width=0.45\textwidth]{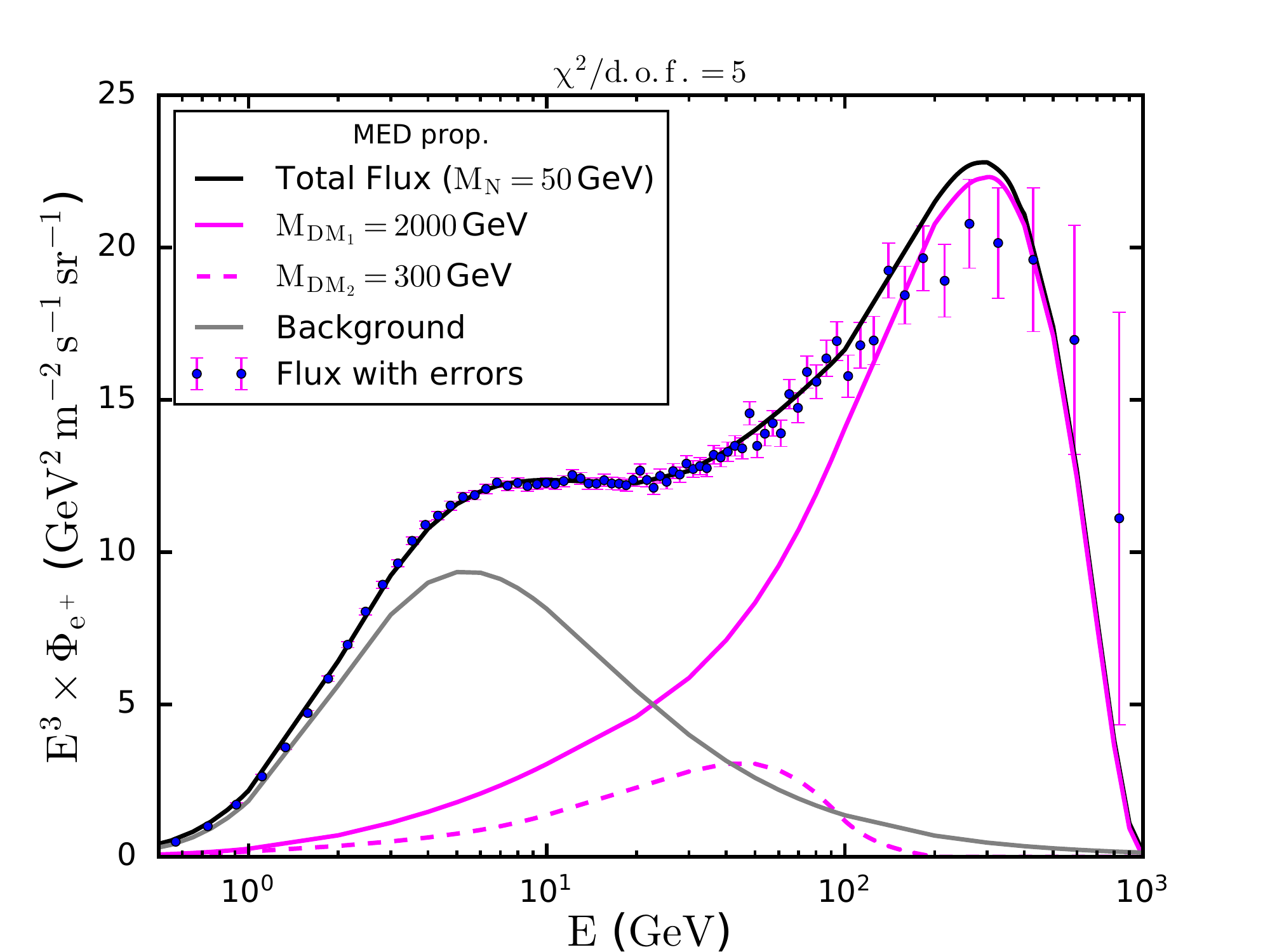}
    \includegraphics[width=0.45\textwidth]{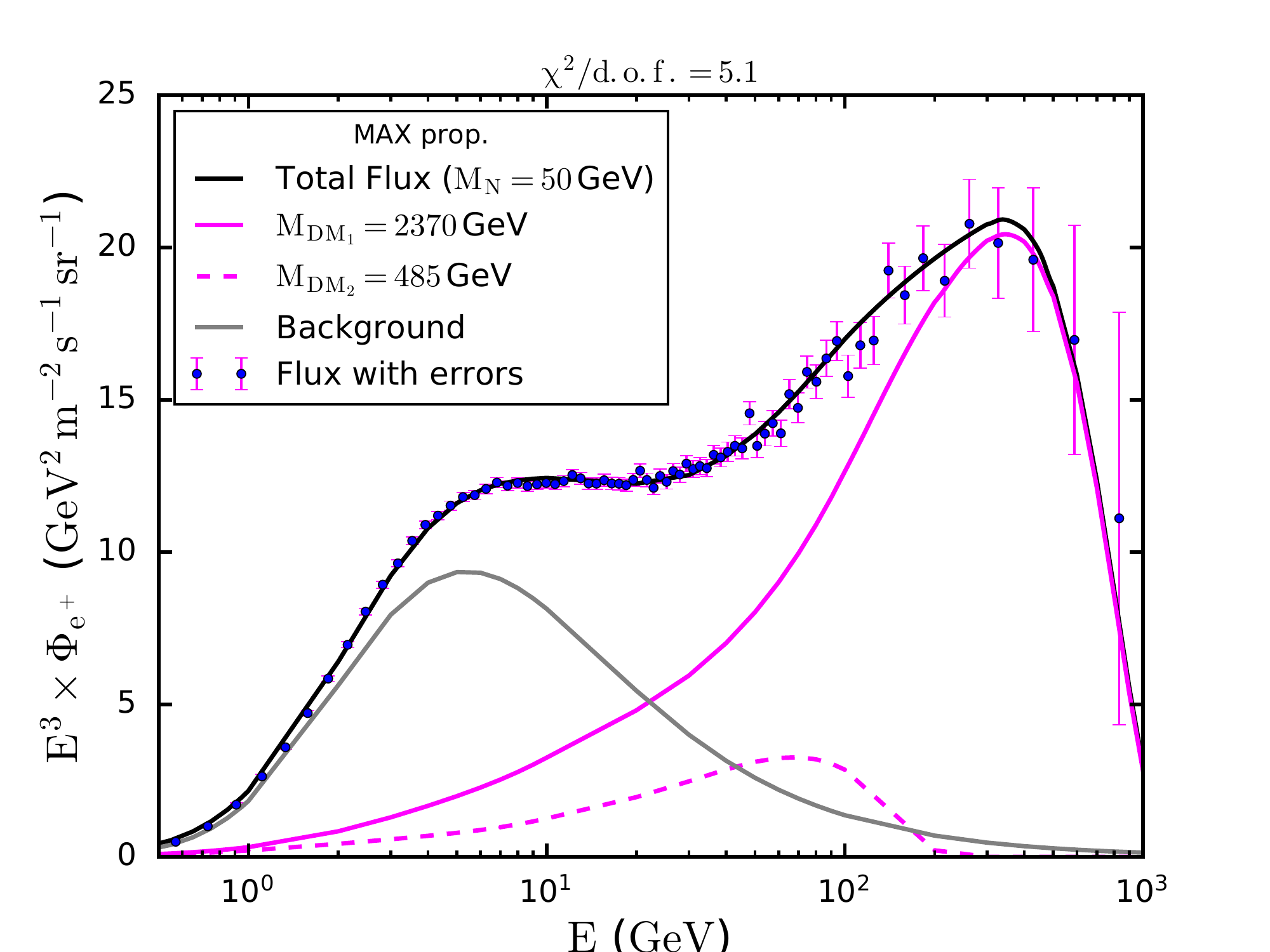}
    \caption{Expected positron flux \textit{versus} energy, summing over different contributions (black line): Dark Matter candidate 1 (continuous magenta line), Dark Matter candidate 2 (dashed magenta line), Diffuse Background (gray line). In this case we choose the right-handed neutrino mass equal to $50$~GeV, for two different propagation models, MED (top) and MAX (bottom), with $\chi^2/d.o.f.=5$ and $\chi^2/d.o.f.=5.1$, respectively.}
    \label{fig:flux50MED}
\end{figure}

To be conservative, in these first analyses, we compute the goodness of the fit, $\chi^2/d.o.f$, using only the statistical uncertainties provided by the collaboration. For each scenario, we chose the best values within $3\sigma$ error for the parameters $c_d$ and $\gamma_d$ to get the best values for the fit, according to the Table~\ref{tab:prop_par}.

\begin{figure}[ht!]
    \centering
    \includegraphics[width=0.45\textwidth]{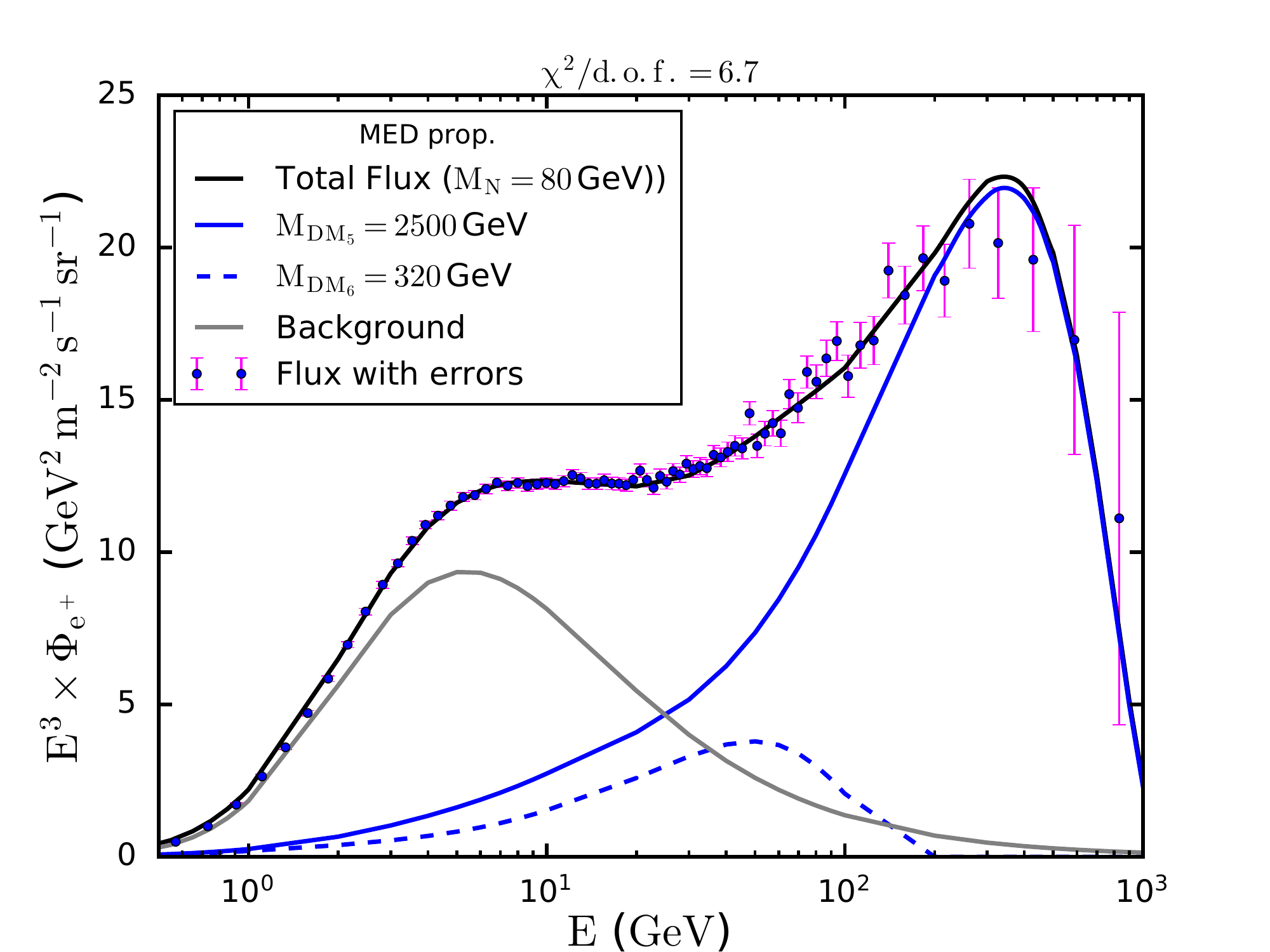}
    \includegraphics[width=0.45\textwidth]{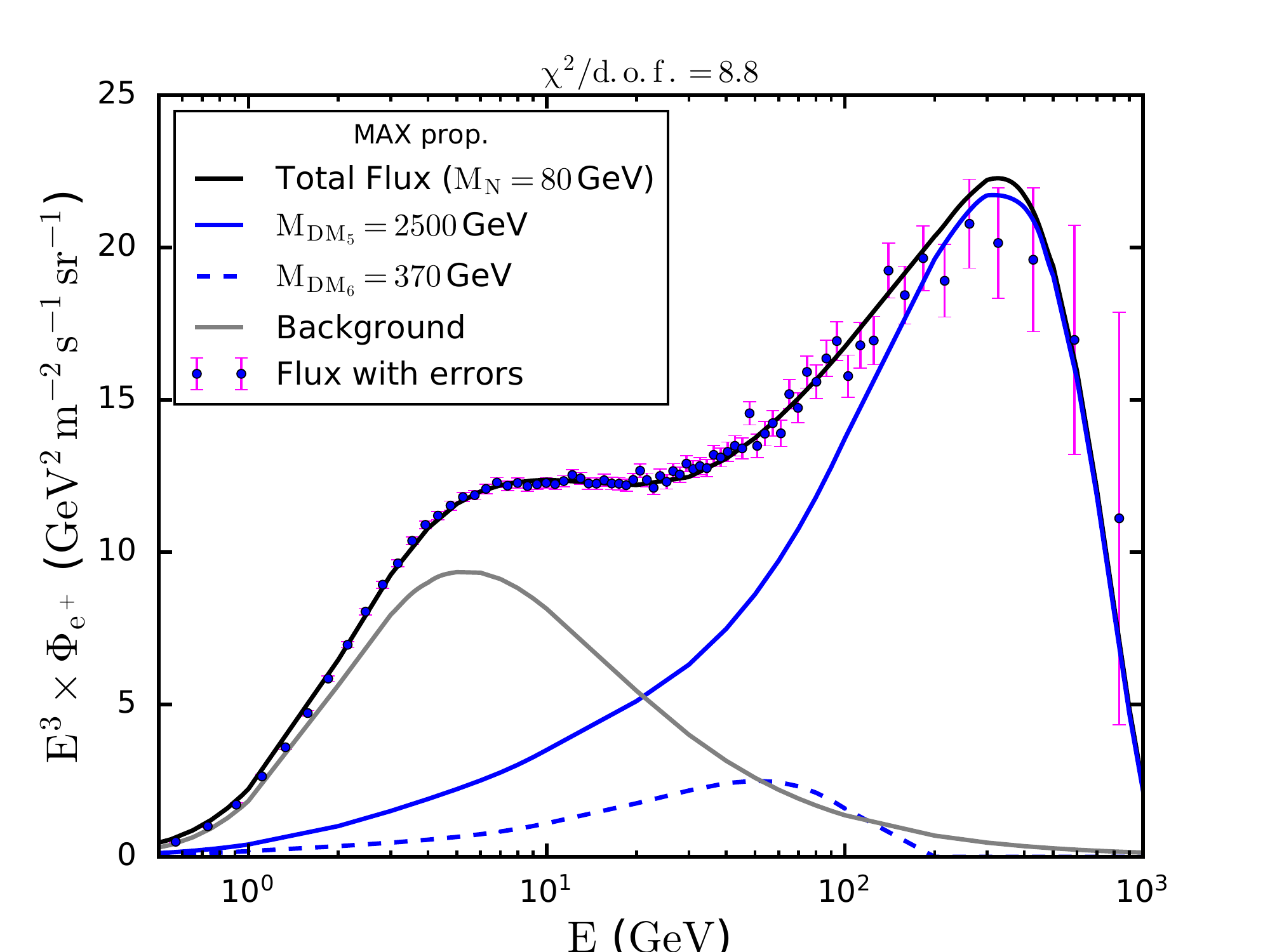}
    \caption{Expected positron flux summing over different contributions (black line): Dark Matter candidate 1 (continuous blue line), Dark Matter candidate 2 (dashed blue line), Diffuse Background (gray line). In this case we choose the right-handed neutrino mass equal to $80$~GeV, for two different propagation models, MED (top) and MAX (bottom), with $\chi^2/d.o.f.=6.7$ and $\chi^2/d.o.f.=8.8$, respectively.}
    \label{fig:flux80MED}
\end{figure}

In Figs.~\ref{fig:flux10MED}, \ref{fig:flux50MED} and \ref{fig:flux80MED}, we present the computed fluxes including that predicted by each decaying DM component (continuous and dashed lines), the background contribution (gray lines) and the sum over all components (black lines). The AMS data \cite{Aguilar:2019owu} is also shown for comparison.   

\begin{figure}[ht!]
    \centering
    \includegraphics[width=0.45\textwidth]{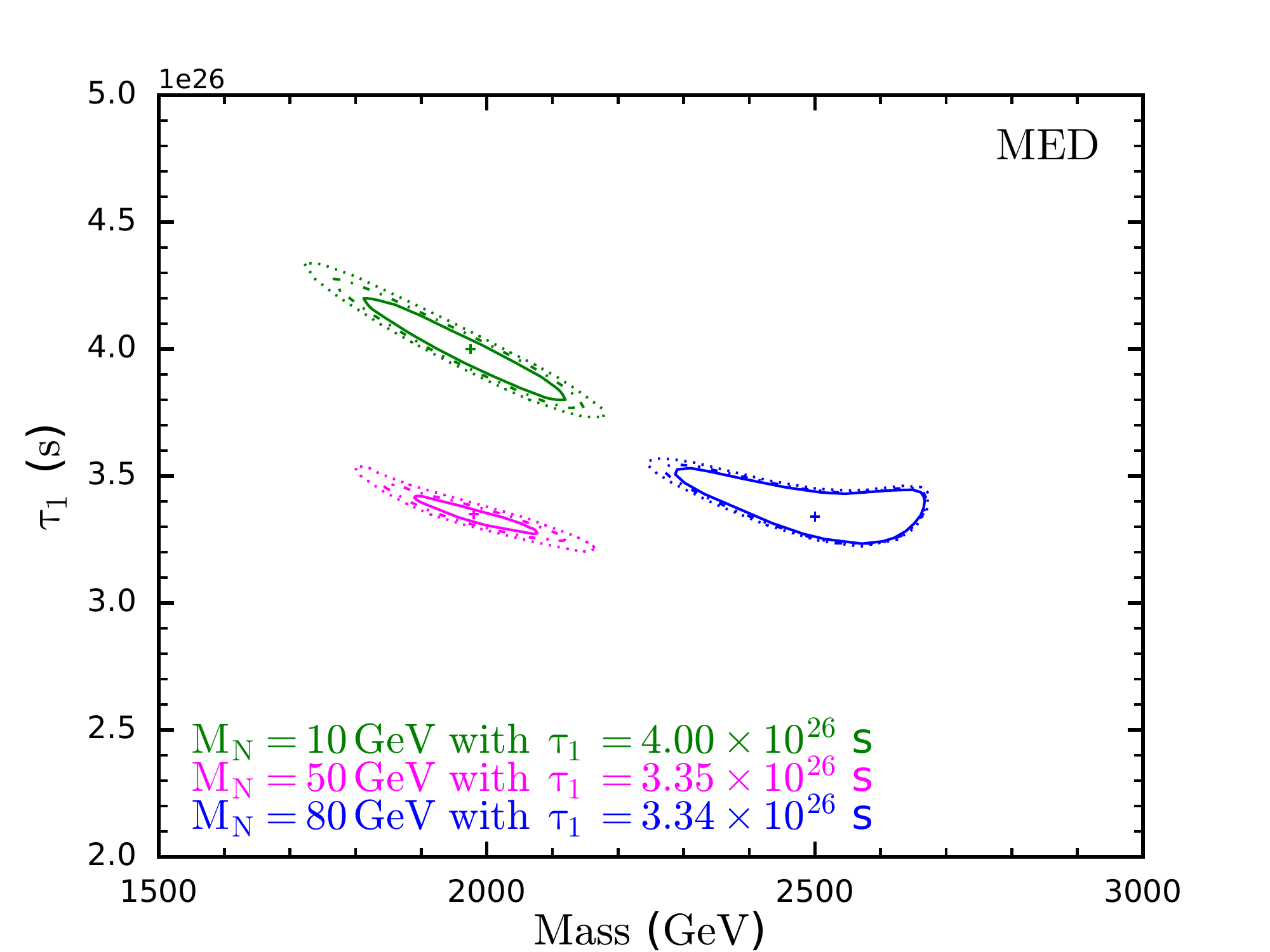}
    \includegraphics[width=0.45\textwidth]{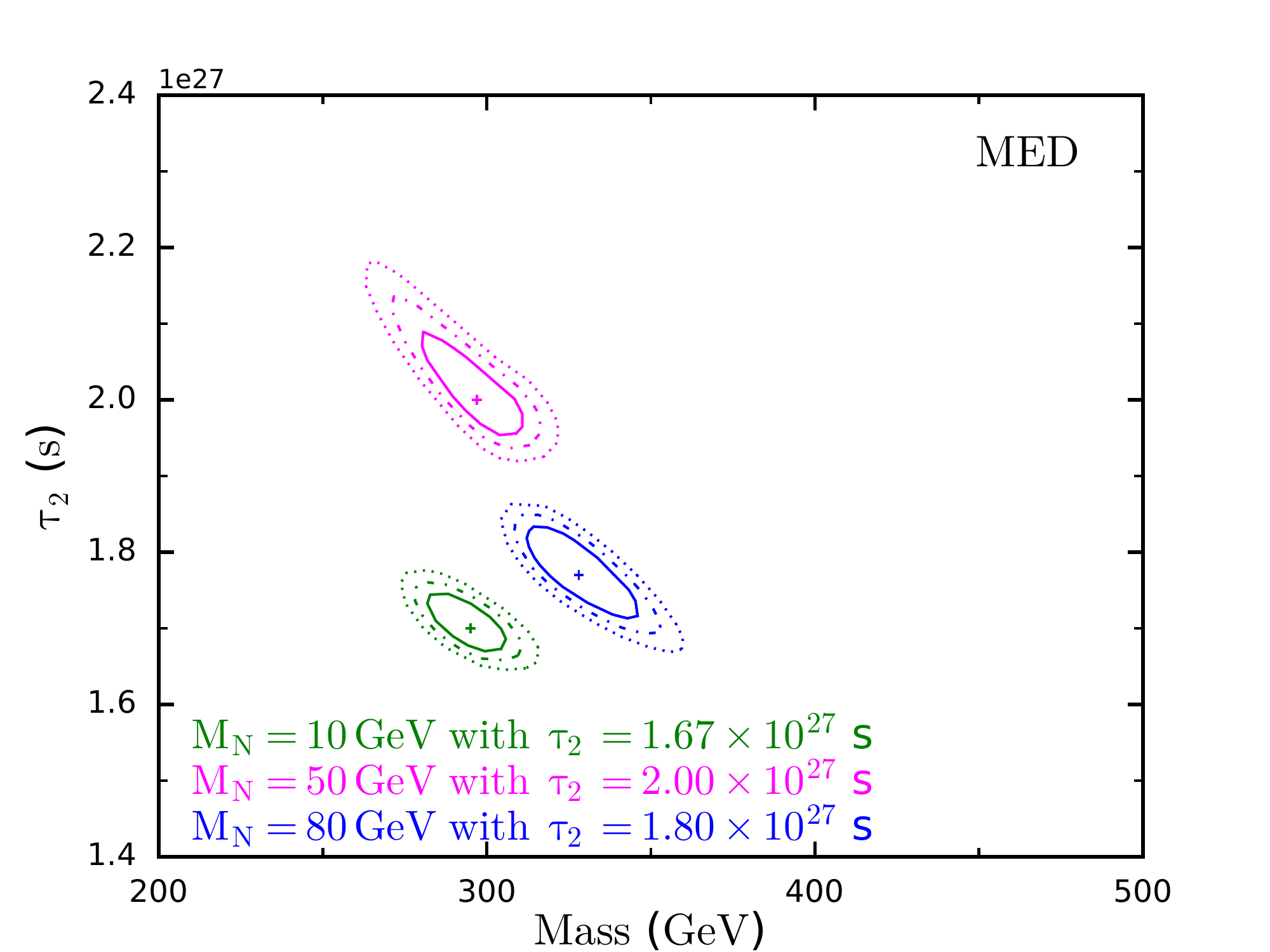}
    \caption{$1\sigma$ (continuous lines), $2\sigma$ (dashed lines) and $3\sigma$ (dotted lines) regions for the Dark Matter candidate 1 (top) and for the Dark Matter candidate 2 (bottom) for different right-handed neutrino masses, for $M_N=10$~GeV (green), $M_N=50$~GeV (magenta), and for $M_N = 80$~GeV (blue).}
    \label{fig:cs1MED}
\end{figure}

\begin{table}[ht]
    \centering
    \begin{tabular}{|c|c|c|c|}
    \hline
    Propag. Model & $M_N$ (GeV) & $c_d$ $({\rm m^2\,sr\,s\,GeV})^{-1}$ & $\gamma_d$ \\
    \hline
                  & $10$ & $6.4\times10^{-2}$ & $-4.02$ \\
       MED        & $50$ & $6.2\times10^{-2}$ & $-4.00$ \\
                  & $80$ & $6.3\times10^{-2}$ & $-4.00$ \\    
    \hline     
                  & $10$ & $6.4\times10^{-2}$ & $-4.02$ \\
       MAX        & $50$ & $6.1\times10^{-2}$ & $-3.97$ \\
                  & $80$ & $6.0\times10^{-2}$ & $-3.98$ \\ 
    \hline
    \end{tabular}
    \caption{Background parameters used in each analysis within $3\sigma$ uncertainties. The values were chosen in order to get the best values for the $\chi^2/d.o.f.$.}
    \label{tab:prop_par}
\end{table}

In Fig.~\ref{fig:flux10MED}, we show our results for $M_N=10$~GeV for two different propagation models, MED and MAX (top and bottom, respectively). We found the best fit value equal to $\chi^2/d.o.f.=3.57$ for the MED propagation model and $\chi^2/d.o.f.=2.3$ for the MAX propagation. In Table \ref{tab:prop_par}, we include the background parameters $c_d$ and $\gamma_d$ adopted for each scenario. 

\begin{figure}[ht!]
    \centering
    \includegraphics[width=0.45\textwidth]{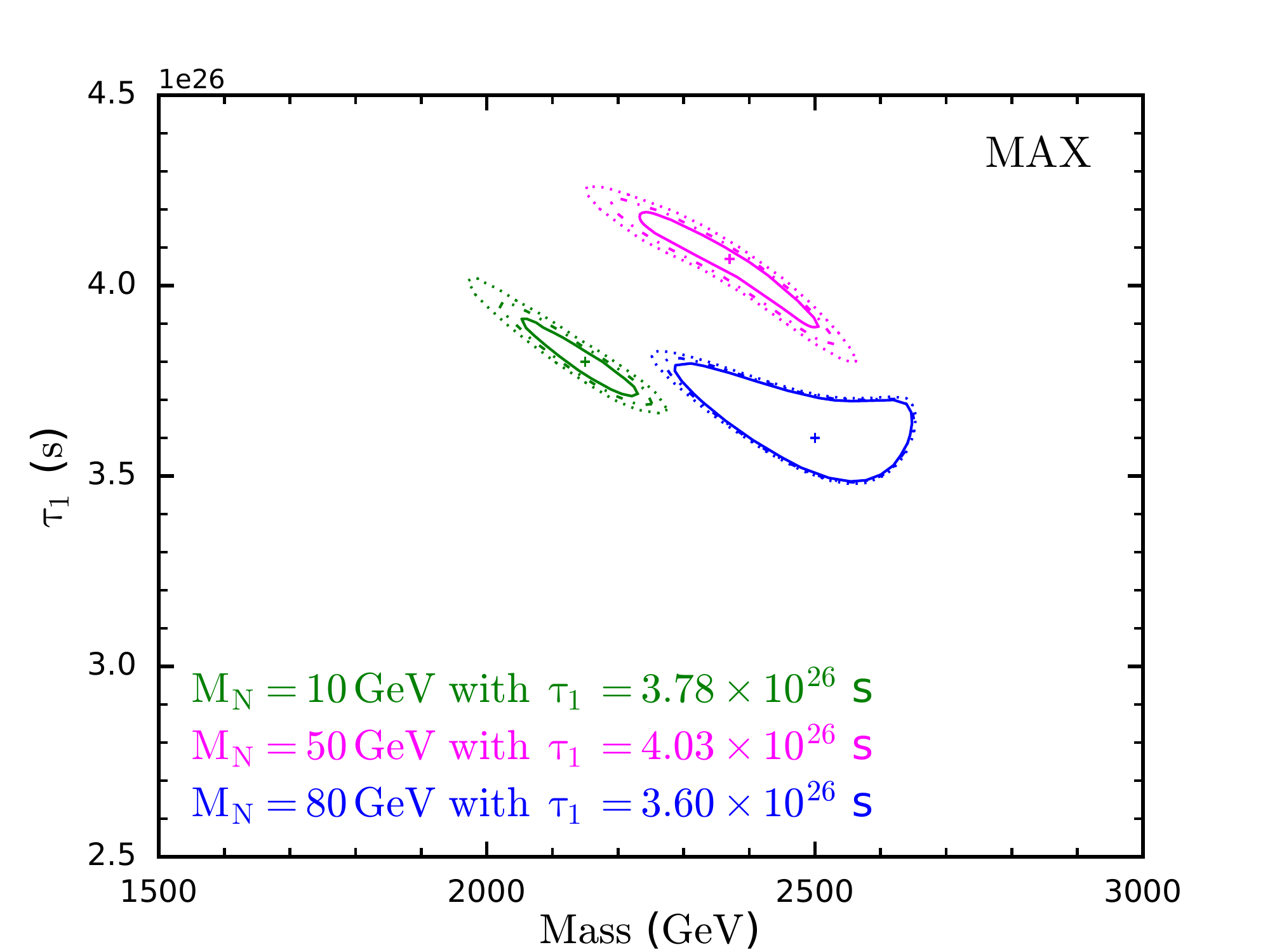}
    \includegraphics[width=0.45\textwidth]{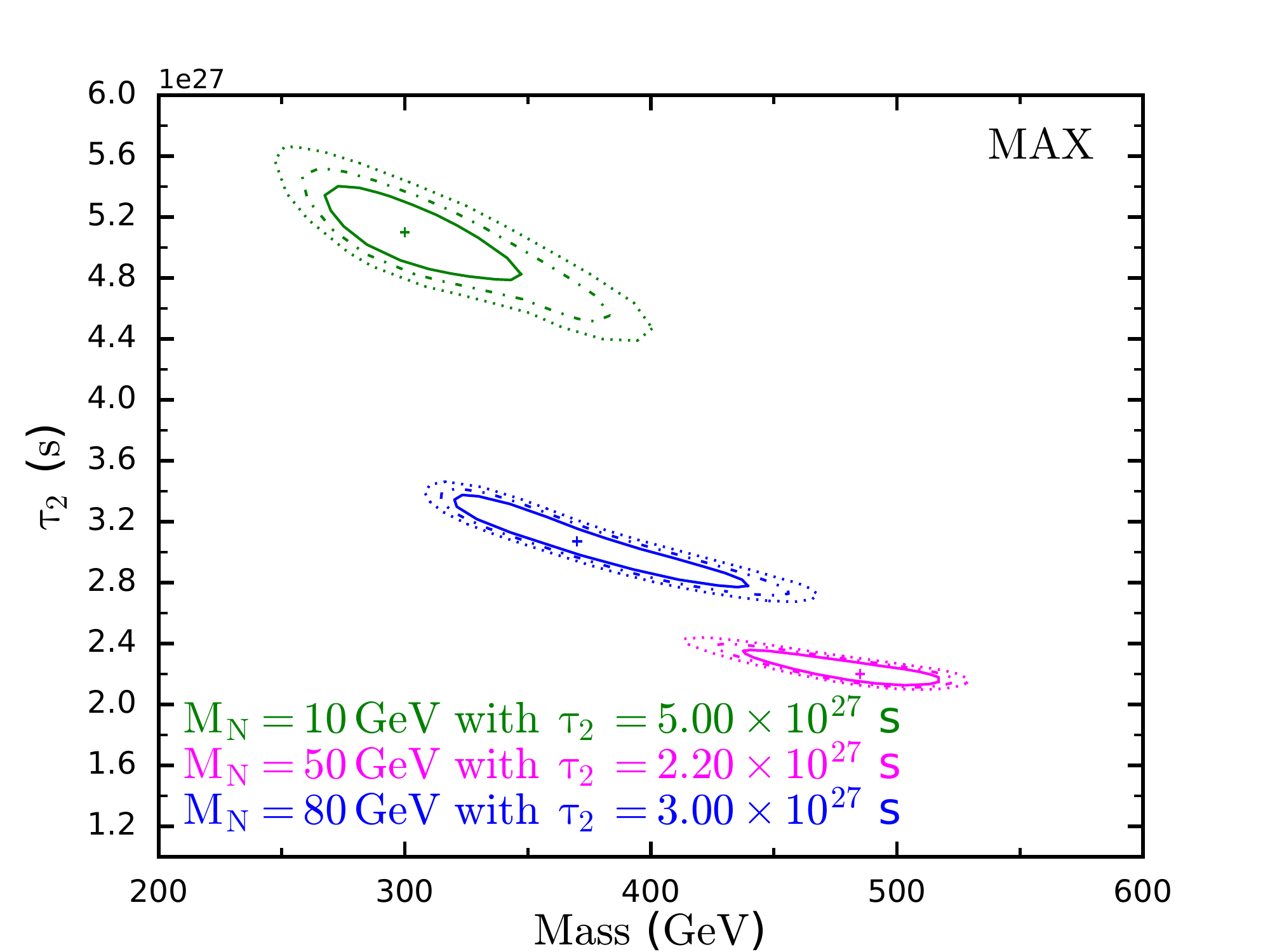}
    \caption{$1\sigma$ (continuous lines), $2\sigma$ (dashed lines) and $3\sigma$ (dotted lines) regions for the Dark Matter candidate 1 (top) and for the Dark Matter candidate 2 (bottom) with different right-handed neutrino masses, for $M_N=10$~GeV (green), $M_N=50$~GeV (magenta), and for $M_N = 80$~GeV (blue).}
    \label{fig:cs1MAX}
\end{figure}

In Fig.~\ref{fig:flux50MED}, we show our results taking $M_N=50$~GeV for two different propagation models, MED and MAX, following the same description above. We found the best fit value equal to $\chi^2/d.o.f.=5.0$ for the MED propagation model and $\chi^2/d.o.f.=5.1$ for the MAX propagation. For $M_N=80$~GeV, we found $\chi^2/d.o.f.=6.7 \,\,(8.8)$ for MED (MAX) propagation (Fig.~\ref{fig:flux80MED}). 

\begin{table*}[ht]
    \centering
    \begin{tabular}{|c|c|c|c|c|c|c|}
    \hline
    Propag. Model & $M_N$ (GeV) & $M_{DM_1}$ (GeV) & $\tau_{1}$ (s) & $M_{DM_2}$ (GeV) & $\tau_{2}$ (s) & $\chi^2/d.o.f.$\\
    \hline
                  & $10$ & $2000$ & $4.00 \times 10^{26}$ & $300$ & $1.67 \times 10^{27}$ & $3.57$\\
       MED        & $50$ & $2000$ & $3.35 \times 10^{26}$ & $300$ & $2.00      \times 10^{27}$ & $5.0$\\
                  & $80$ & $2500$ & $3.34 \times 10^{26}$ & $320$ & $1.80 \times 10^{27}$ & $6.7$\\    
    \hline     
                  & $10$ & $2150$ & $3.78 \times 10^{26}$ & $300$ & $5.00 \times 10^{27}$ & $2.3$ \\
       MAX        & $50$ & $2370$ & $4.07 \times 10^{26}$ & $485$ & $2.20 \times 10^{27}$ & $5.1$\\
                  & $80$ & $2500$ & $3.60 \times 10^{26}$ & $370$ & $3.00 \times 10^{27}$ & $8.8$\\ 
    \hline
    \end{tabular}
    \caption{Best-fit parameters found for the different scenarios.}
    \label{tab:best_par}
\end{table*}

Summarizing, in Table~\ref{tab:best_par} we show the best-fit values found for the parameters DM mass and lifetime for each DM candidate in order to get the best fit to the data. In Figs.~\ref{fig:cs1MED} (MED propagation) and \ref{fig:cs1MAX} (MAX propagation), we present the $1\sigma$, $2\sigma$ and $3\sigma$ contours for both DM1 (top) and DM2 (bottom) candidates following the same color pattern for the RHN masses described above.

As we can see, the larger the right handed neutrino mass the worst is the fit to data. This is due to the change in the shape of the spectrum. Although their shapes seems to be quite similar, minimum modifications in the tale (lower energies) provide a significant impact on the $\chi^2/d.o.f.$ as a result of the smallness of the error bars at lower energies.

In the same way, the MED propagation model yields smaller fluxes than MAX propagation one. Hence, we can play with the decay rate (or lifetime) in order to obtain similar fits for both propagation models. For example, taking $M_N=10$~GeV  the best fit is found for $\tau_2=1.67 \times 10^{27}$~s and $M_{DM_2}=300$~GeV for MED propagation while for MAX we need $5.0\times 10^{27}$~s (see Table~\ref{tab:best_par}). As the MAX models gives rise to the steeper energy spectrum, we need to increase the lifetime to find a similar fit.

The combination of two different candidates can provide an excellent agreement with AMS excess, including one of them with mass around hundreds of GeV and another with mass of a few TeV. It is worth emphasizing that the choice $50\%$-$50\%$ for each DM candidate is arbitrary, in a way that modification of this percentage results simply in a re-scaling of the lifetime. 

\subsection{Including Systematic Uncertainties}

The previous analysis included just statistical uncertainties which can be considered conservative as the interpretation of the AMS data is dominated by systematics. Here, we include systematic uncertainties in order to verify its impact in the limits. We concluded that the main impact occurs at lower energies which features rather small error bars. Therefore, the impact in the $\chi^2/d.o.f.$ can be large but usually decreases by a factor of a few. In our study, we choose the MED propagation, with systematic uncertainties provided by the collaboration \cite{Aguilar:2019owu}. One can easily realize that the choice for {\it MED} or {\it MAX} propagation model does not result in significant changes to our conclusions and for this reason we picked the focused on the {\it MED} model in this particular analysis. We emphasize that our conclusions would still apply for the {\it MAX} propagation model. We repeat the procedure above and assume that each dark matter particle contributes to $50\%$ of the dark matter density.

\begin{figure}[ht]
    \centering
    \includegraphics[scale=0.45]{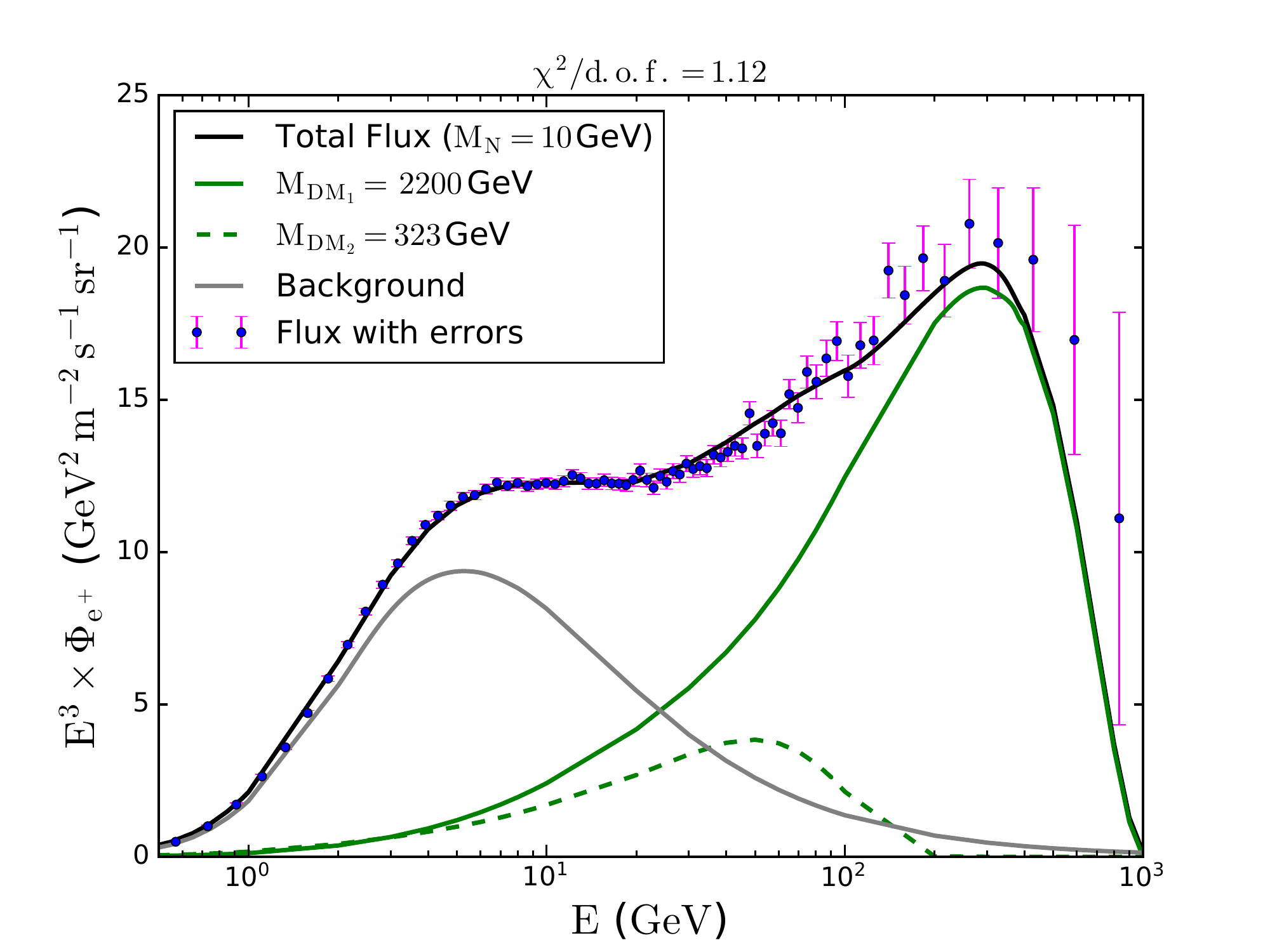}
    \caption{Expected positron flux \textit{versus} energy, summing over different contributions (black line): Dark Matter candidate 1 (dashed line), Dark Matter candidate 2 (continuous line), Diffuse Background (gray line). In this case we choose the right-handed neutrino mass equal to $10$~GeV, and we found $\chi^2/d.o.f.=1.12$.}
    \label{fig:flux10}
\end{figure}

\begin{figure}[ht]
    \centering
    \includegraphics[scale=0.45]{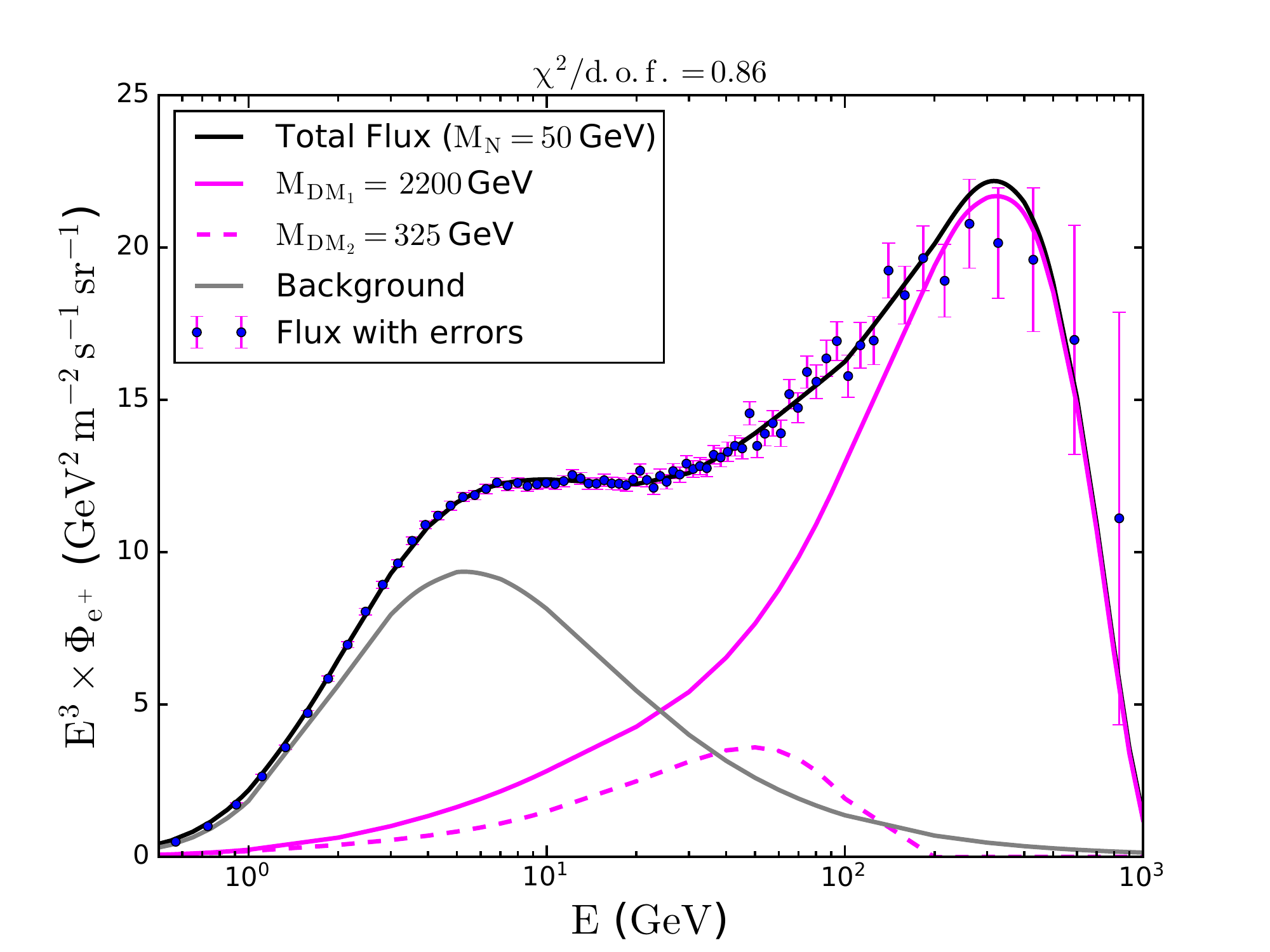}
    \caption{Expected positron flux \textit{versus} energy, summing over different contributions (black line): Dark Matter candidate 1 (dashed line), Dark Matter candidate 2 (continuous line), Diffuse Background (gray line). In this case we choose the right-handed neutrino mass equal to $50$~GeV, and we found $\chi^2/d.o.f.=0.86$.}
    \label{fig:flux50}
\end{figure}

In the Fig.~\ref{fig:flux10}, we present the fluxes that yield the best-fit for  $M_N=10$~GeV. As shown in the Fig.~\ref{fig:flux10} we obtained $M_{DM_1}=2200$~GeV (green continuous line) and $M_{DM_2}=323$~GeV (green dashed line). The diffuse flux (gray line) and total flux (black line) as also exhibited. This setup results in $\chi^2/d.o.f.=1.12$. 

\begin{figure}[ht]
    \centering
    \includegraphics[scale=0.45]{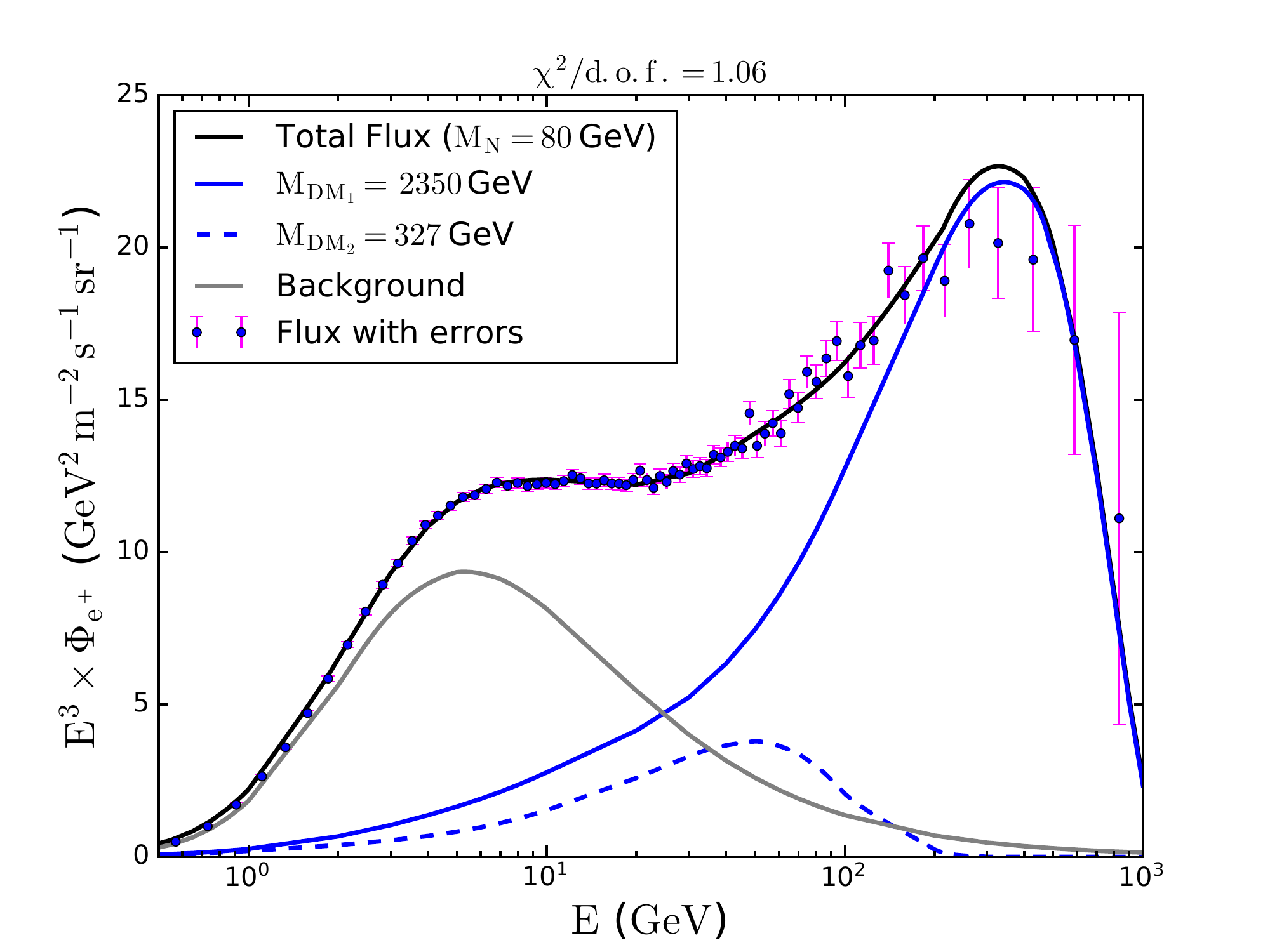}
    \caption{Expected positron flux \textit{versus} energy, summing over different contributions (black line): Dark Matter candidate 1 (solid line), Dark Matter candidate 2 (dashed line), Diffuse Background (gray line). In this case we choose the right-handed neutrino mass equal to $80$~GeV, and we found $\chi^2/d.o.f.=1.06$.}
    \label{fig:flux80}
\end{figure}

In the Fig.~\ref{fig:flux50}, we exhibit our results assuming $M_N=50$~GeV. The best-fit point  yields $\chi^2/d.o.f.=0.86$ and it  is found for $M_{DM_1}=2200$~GeV (pink continuous line) and $M_{DM_2}=323$~GeV (pink dashed line). 

\begin{figure}[ht]
    \centering
    \includegraphics[width=0.45\textwidth]{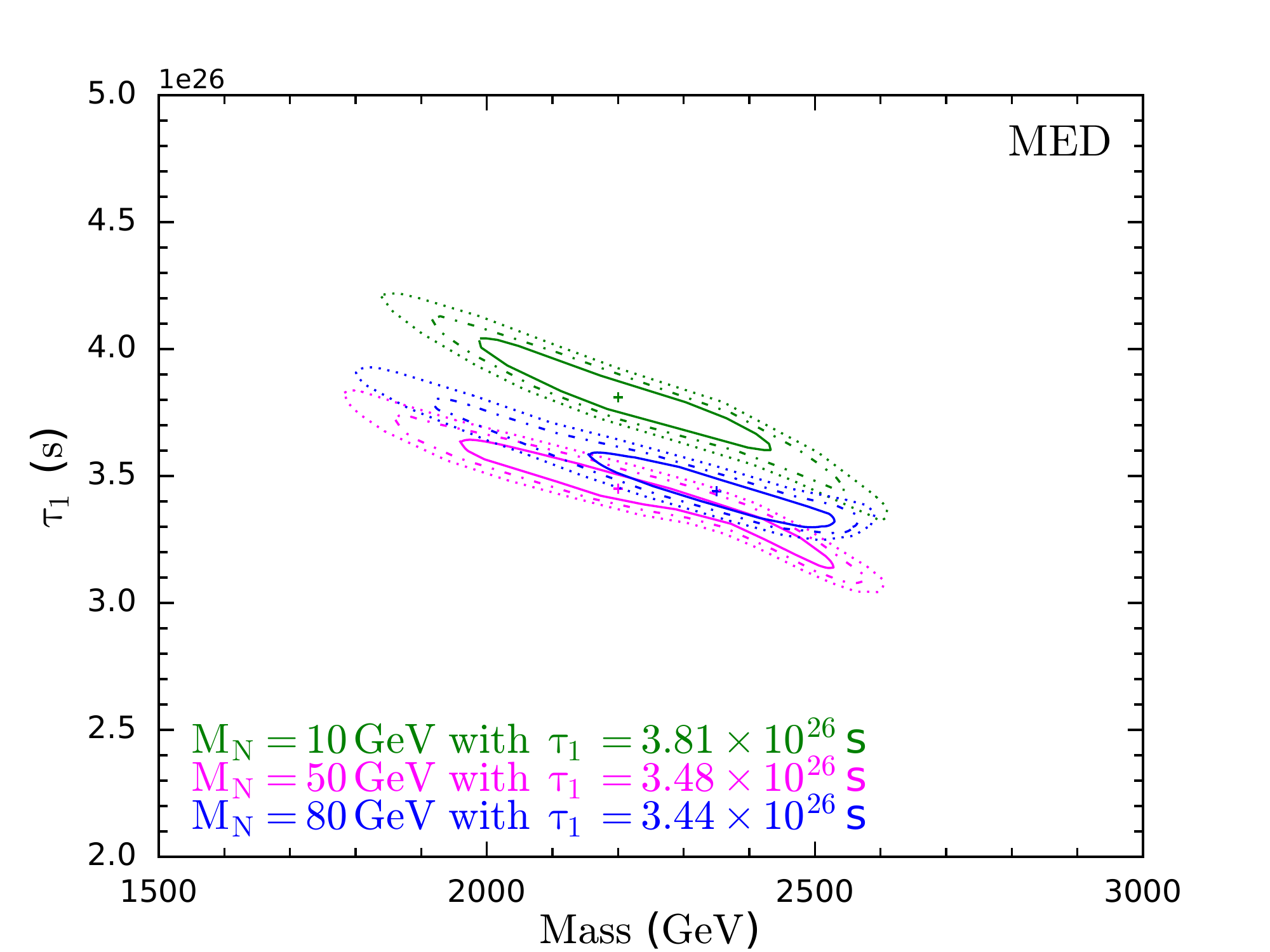}
    \includegraphics[width=0.45\textwidth]{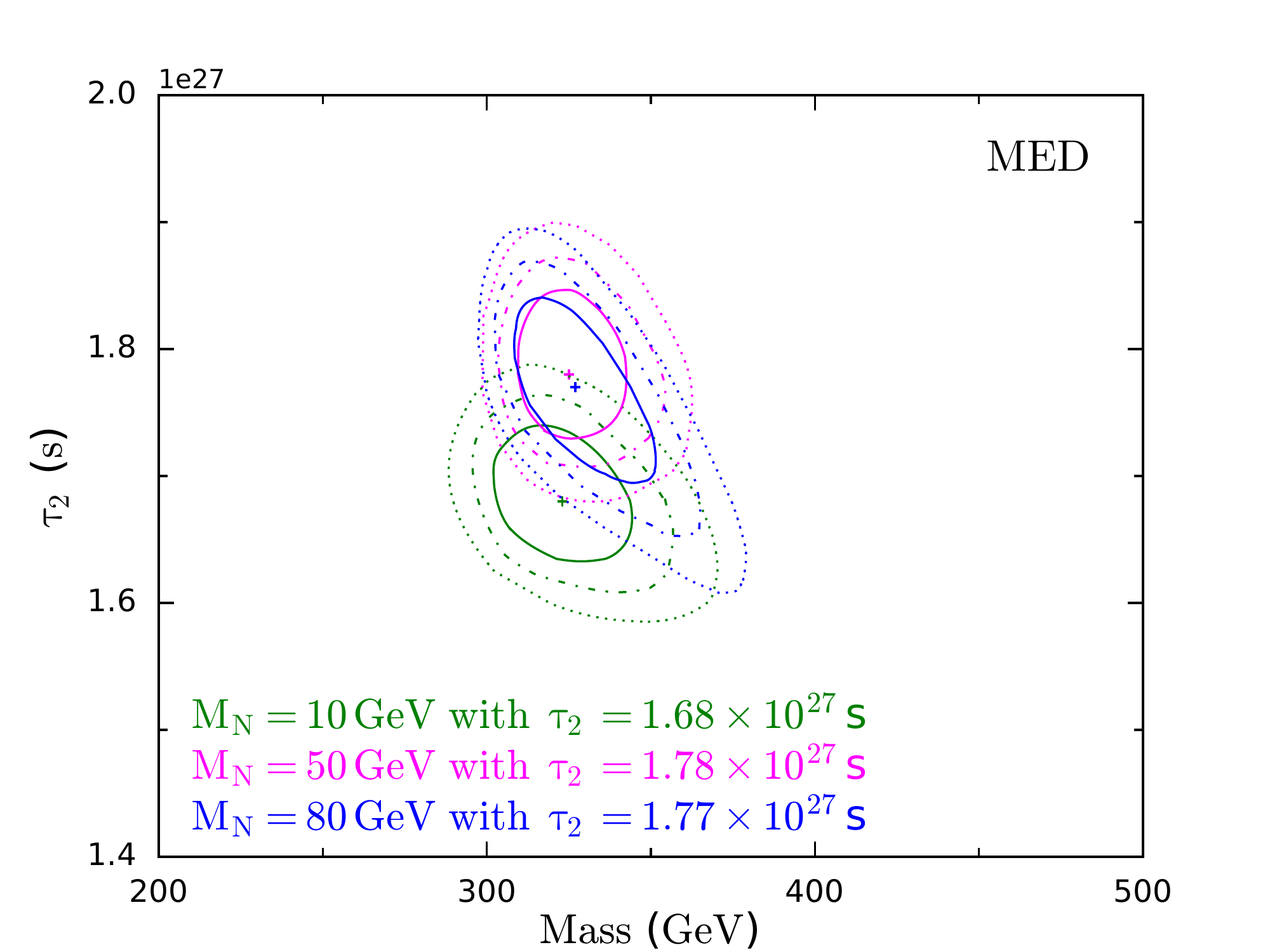}
    \caption{$1\sigma$ (continuous lines), $2\sigma$ (dashed lines) and $3\sigma$ (dotted lines) regions for the Dark Matter candidate 1 (top) and for the Dark Matter candidate 2 (bottom) with different right-handed neutrino masses, for $M_N=10$~GeV (green), $M_N=50$~GeV (magenta), and for $M_N = 80$~GeV (blue).}
    \label{fig:cs1}
\end{figure}

In the Fig.~\ref{fig:flux80}, we repeat the exercise for  $M_N=80$~GeV which still provides a good fit to date with $\chi^2/d.o.f.=1.06$ for $M_{DM_1}=2350$~GeV (blue continuous line) and $M_{DM_2}=327$~GeV (blue dashed line). 

We have explicitly shown that our benchmark scenarios provide a good fit to data and now display the best-fit contours ($1\sigma$ (continuous lines), $2\sigma$ (dashed lines) and $3\sigma$ (dotted lines) contours) in terms of the lifetime and dark matter mass in the Fig.~\ref{fig:cs1} for each setup discussed where both statistical and systematic errors are included.

One could find that the best-fit is found for $M_N=50$~GeV, however, since the AMS data is driven by systematic errors is reasonable to conclude that all of them provide an equally good fit to data. For concreteness, we computed the p-value and we found, for instance, including systematic uncertainties, $p-value =0.25$ for $M_N=10$~GeV, $p-value =0.78$ for $M_N=50$~GeV,  and $p-value =0.33$ for $M_N=80$~GeV. Thus, we did find a good fit to the data. However, we would like to stress that the statistical method used is not of utmost importance because the AMS-02 data is driven by systematic.

\section{Discussion}

The two component dark matter scenario where a scalar (or vector) decays into right-handed neutrino pair was motivated by scalar models which embed the type I seesaw mechanism. In the type I seesaw mechanism the right-handed neutrinos are typically very heavy, however we found that for masses heavier than $\sim 100$~GeV the fit to the AMS data becomes quite poor. This can be understood via the energy spectrum. When right-handed neutrinos are heavier than $100$~GeV, the decay channels into Z and W bosons are open leading to significant changes in the energy spectrum, and as we checked, it provides a poor fit to data. That said, even in the type I seesaw mechanism we can easily assume right-handed neutrino masses between 10-80 GeV by tuning the Yukawa couplings, bringing no changes to the branching ratio pattern, which justifies our analysis.

Another aspect of our study is the compatibility with limits stemming from gamma-ray data, because our decay channels also produce gamma-rays. Our setup involved dark matter decaying into right-handed neutrino pairs where each right-handed neutrino might decay into leptons and quarks via off-shell W, Higgs and Z bosons. Thus, as we have not fixed a final decay channel it is not so simple to compare out with other existing limits in the literature. Sifting the energy spectra produced by DM decay into SM particles, we realized that the gamma-ray spectrum produced by a direct DM decay into $WW$ and $W\ell$ though different, yield the closest shape to the energy spectra produced by our setup. Thus we can compare the energy spectra and notice by how much different they are, and then re-scale our energy spectra by a given amount to match the energy spectra of the $WW$ and $W\ell$ channels. In this way, we may roughly estimate whether our benchmark points are in agreement with existing gamma-ray limits \cite{Cohen:2016uyg}. We concluded that taking into account the facts that we have a two component dark matter setup and the uncertainties involved in the gamma-ray limits our benchmark points are consistent with the existing gamma-ray bounds. Although, we highlight that there are no existing gamma-ray limits directly applicable to our model, and that required an extra effort from our side to somehow compare our results with gamma-ray probes that feature a similar energy spectrum. We will prolong this discussion in the {\it Appendix}. 

In summary, we have shown that such two component dark matter via the right-handed neutrino portal offer a good fit to data for right-handed neutrino masses between 10-80 GeV with the inclusion or not of systematic errors in the analysis. Within this mass range, the precise mass of the right-handed neutrino does not change much the lifetime and dark matter mass that best fit the data, but do change the $\chi^2/d.o.f$ by a factor of two. In addition to that, the change from {\it MED} to {\it MAX} propagation model does not bring significant changes to our study, despite the {\it MAX} propagation being recently favored by recent observations of the Boron-to-Carbon ratio \cite{Genolini:2019ewc}. In our study we concluded that masses around $300$~GeV and $2$~TeV with lifetime of $4\times 10^{26}$~s and $2 \times 10^{27}$~s respectively, are favored and marginally consistent with current bounds rising from gamma-ray observations. 
 
\section{Conclusions}
\label{sec:con}

The positron excess provided by the AMS collaboration \cite{Aguilar:2019owu} remains an open question. In this work we assessed a scenario where two decaying dark matter candidates may constitute an answer to the observed excess via the right-handed neutrino portal.

We have shown that DM particles decaying into right handed neutrino pairs which couples to SM particles through $Z$, $W$ and Higgs bosons, inspired by the type I seesaw mechanism provide a very good fit to data. For example, for a conservative approach including just statistical uncertainties we got $\chi^2/d.o.f \sim 2.3$ for $m_{DM_1}=2150$~GeV with $\tau_{1}=3.78 \times 10^{26}$~s and $m_{DM_2}=300$ with $\tau_{2}=5.0 \times 10^{27}$~s for $M_N=10$~GeV, and, in an optimistic case, including systematic uncertainties, we found $\chi^2/d.o.f \sim 1.12$, for $M_N = 10$~GeV, with $m_{DM_1}=2200$~GeV with $\tau_{1}=3.8 \times 10^{26}$~s and $m_{DM_2}=323$~GeV with $\tau_{2}=1.68 \times 10^{27}$~s. 

Such benchmark points are consistent with existing gamma-ray bounds for lighter DM but in tension for heavier DM, however, as described in the appendix, due to the large uncertainties in the gamma-ray limits, we may argue that our benchmarks are in agreement with gamma-ray data. It is important to emphasize that this is an estimate, and a careful analysis is needed.  In addition, our benchmarks are significantly modified by changing the propagation model from MED to MAX. Knowing that the AMS results are dominated by systematics our best-fit points might alter for different assumption for the background. In our work we adopted the background recommended by the AMS collaboration. 

In summary, we presented a plausible explanation to the puzzling AMS data via the right-handed neutrino portal.

\section*{Acknowledgments}
The authors thank Pasquale Serpico, Diego Restrepo, Manuela Vecchi and Joseph Silk for useful discussions. FSQ acknowledges support from CNPq grants 303817/2018-6 and 421952/2018-0, UFRN, MEC and ICTPSAIFR FAPESP grant 2016/01343-7. CS thanks UFRN and MEC for the financial support. We thank the High Performance Computing Center (NPAD) at UFRN for providing computational resources.

\section{Appendix}

\begin{figure*}[ht]
    \centering
    \includegraphics[width=0.95\textwidth]{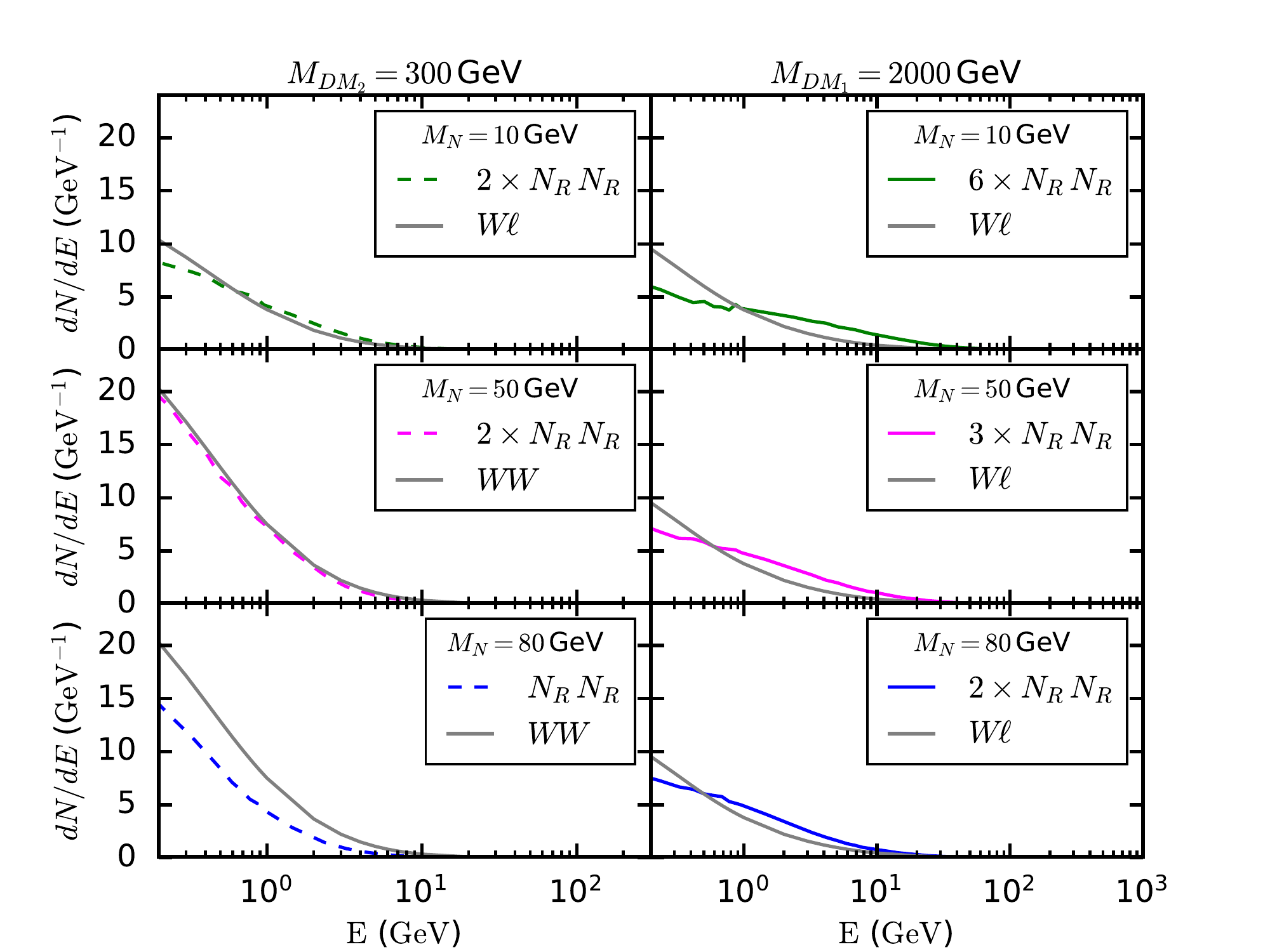}
    \caption{Comparison between the gamma-ray spectrum provided by direct annihilations into $WW$ and $W\ell$ with the spectrum provided by our right handed neutrinos re-scaled accordingly. For details please see the text.}
    \label{comparison}
\end{figure*}

The decay into right-handed neutrinos also produces gamma-rays, thus we need to check if our scenario is in agreement with existing gamma-ray observations. Although, there is no gamma-ray limit in the literature for dark matter decaying into right-handed neutrinos. Thus in order to estimate if our best-fit points are then consistent gamma-ray bounds we looked after the popular decay channels to check which ones produce similar gamma-ray spectra. They are all different, but the ones that resemble most our case are the decay into WW and $W\ell$. To explicitly show our procedure we chose a benchmark scenario with $M_N=10,\,50,\,80$~GeV, where the best fit to the positron data is given by $M_{DM_1}\simeq 2000$~GeV and $M_{DM_2}=300$~GeV. In order to get a comparable limit, we need to re-scale the spectrum according to the Fig.~\ref{comparison} below. We highlight that the $WW$ and $W\ell$ spectra were re-scale by different a constant factor to approximate their spectra to ours.  For example, for $M_{DM_1}\simeq 2000$~GeV and $M_N=10$~GeV we had to multiply the our spectrum by six. Therefore, the limits provided by \cite{Cohen:2016uyg} need to be suppressed also by a factor of six times to be applicable to our setup. Moreover, an additional  factor $1/2$ should be included due to the DM density since our case we have DM components. That said, the gamma-ray limit from \cite{Cohen:2016uyg} at face value reads $3.6 \times 10^{28}$~s, but it should be read as $3 \times 10^{27}$~s. While for $M_{DM_2}=300$~GeV, at face value the limit reads $4.8 \times 10^{27}$~s \cite{Cohen:2016uyg}, but taking into account the factors provides $1.2 \times 10^{27}$~s. 

Using these estimates, we conclude that the lighter DM candidate is in agreement with the limits while the heavier not by a factor of a few. Having in mind that the limits in \cite{Cohen:2016uyg} are optimistic due to the profile and target selected (inner galaxy) these gamma-ray bounds are subject to large uncertainties, we may argue that our best fit points are marginally in agreement with the gamma-ray bounds. A similar reasoning could be applied to different benchmark points.

In addition, we emphasize that any assessment of the best-fit points rely on the background model assumed for the positron secondary production resulted from the collision of primary cosmic rays with the interstellar medium. In our work, we adopt the background model used by AMS-02 collaboration in their data release \cite{Aguilar:2019owu}, thus our conclusions are based on that. There are other competitive gamma-ray bounds in the literature \cite{Ando:2016ang} which can be also relaxed in a similar way.

\bibliographystyle{JHEPfixed}
\bibliography{darkmatter}

\providecommand{\href}[2]{#2}\begingroup\raggedright\begin{thebibliography}{10}

\bibitem{Adriani:2008zr}
{\bf PAMELA} Collaboration, O.~Adriani {\em et.~al.}, {\it {An anomalous
  positron abundance in cosmic rays with energies 1.5-100 GeV}},  {\em Nature}
  {\bf 458} (2009) 607--609, [\href{http://xxx.lanl.gov/abs/0810.4995}{{\tt
  0810.4995}}].

\bibitem{FermiLAT:2011ab}
{\bf Fermi-LAT} Collaboration, M.~Ackermann {\em et.~al.}, {\it {Measurement of
  separate cosmic-ray electron and positron spectra with the Fermi Large Area
  Telescope}},  {\em Phys. Rev. Lett.} {\bf 108} (2012) 011103,
  [\href{http://xxx.lanl.gov/abs/1109.0521}{{\tt 1109.0521}}].

\bibitem{Aguilar:2013qda}
{\bf AMS} Collaboration, M.~Aguilar {\em et.~al.}, {\it {First Result from the
  Alpha Magnetic Spectrometer on the International Space Station: Precision
  Measurement of the Positron Fraction in Primary Cosmic Rays of 0.5–350
  GeV}},  {\em Phys. Rev. Lett.} {\bf 110} (2013) 141102.

\bibitem{Cholis:2013psa}
I.~Cholis and D.~Hooper, {\it {Dark Matter and Pulsar Origins of the Rising
  Cosmic Ray Positron Fraction in Light of New Data From AMS}},  {\em Phys.
  Rev.} {\bf D88} (2013) 023013, [\href{http://xxx.lanl.gov/abs/1304.1840}{{\tt
  1304.1840}}].

\bibitem{Ibarra:2013zia}
A.~Ibarra, A.~S. Lamperstorfer, and J.~Silk, {\it {Dark matter annihilations
  and decays after the AMS-02 positron measurements}},  {\em Phys. Rev.} {\bf
  D89} (2014), no.~6 063539, [\href{http://xxx.lanl.gov/abs/1309.2570}{{\tt
  1309.2570}}].

\bibitem{Hooper:2012sr}
D.~Hooper, C.~Kelso, and F.~S. Queiroz, {\it {Stringent and Robust Constraints
  on the Dark Matter Annihilation Cross Section From the Region of the Galactic
  Center}},  {\em Astropart. Phys.} {\bf 46} (2013) 55--70,
  [\href{http://xxx.lanl.gov/abs/1209.3015}{{\tt 1209.3015}}].

\bibitem{Ackermann:2015zua}
{\bf Fermi-LAT} Collaboration, M.~Ackermann {\em et.~al.}, {\it {Searching for
  Dark Matter Annihilation from Milky Way Dwarf Spheroidal Galaxies with Six
  Years of Fermi Large Area Telescope Data}},  {\em Phys. Rev. Lett.} {\bf 115}
  (2015), no.~23 231301, [\href{http://xxx.lanl.gov/abs/1503.02641}{{\tt
  1503.02641}}].

\bibitem{Galli:2011rz}
S.~Galli, F.~Iocco, G.~Bertone, and A.~Melchiorri, {\it {Updated CMB
  constraints on Dark Matter annihilation cross-sections}},  {\em Phys. Rev.}
  {\bf D84} (2011) 027302, [\href{http://xxx.lanl.gov/abs/1106.1528}{{\tt
  1106.1528}}].

\bibitem{Weniger:2013hja}
C.~Weniger, P.~D. Serpico, F.~Iocco, and G.~Bertone, {\it {CMB bounds on dark
  matter annihilation: Nucleon energy-losses after recombination}},  {\em Phys.
  Rev.} {\bf D87} (2013), no.~12 123008,
  [\href{http://xxx.lanl.gov/abs/1303.0942}{{\tt 1303.0942}}].

\bibitem{Nardi:2008ix}
E.~Nardi, F.~Sannino, and A.~Strumia, {\it {Decaying Dark Matter can explain
  the e+- excesses}},  {\em JCAP} {\bf 0901} (2009) 043,
  [\href{http://xxx.lanl.gov/abs/0811.4153}{{\tt 0811.4153}}].

\bibitem{Arvanitaki:2008hq}
A.~Arvanitaki, S.~Dimopoulos, S.~Dubovsky, P.~W. Graham, R.~Harnik, and
  S.~Rajendran, {\it {Astrophysical Probes of Unification}},  {\em Phys. Rev.}
  {\bf D79} (2009) 105022, [\href{http://xxx.lanl.gov/abs/0812.2075}{{\tt
  0812.2075}}].

\bibitem{Dienes:2013xff}
K.~R. Dienes, J.~Kumar, and B.~Thomas, {\it {Dynamical Dark Matter and the
  positron excess in light of AMS results}},  {\em Phys. Rev.} {\bf D88}
  (2013), no.~10 103509, [\href{http://xxx.lanl.gov/abs/1306.2959}{{\tt
  1306.2959}}].

\bibitem{Geng:2013nda}
C.-Q. Geng, D.~Huang, and L.-H. Tsai, {\it {Imprint of multicomponent dark
  matter on AMS-02}},  {\em Phys. Rev.} {\bf D89} (2014), no.~5 055021,
  [\href{http://xxx.lanl.gov/abs/1312.0366}{{\tt 1312.0366}}].

\bibitem{Belotsky:2014haa}
K.~Belotsky, M.~Khlopov, C.~Kouvaris, and M.~Laletin, {\it {Decaying Dark Atom
  constituents and cosmic positron excess}},  {\em Adv. High Energy Phys.} {\bf
  2014} (2014) 214258, [\href{http://xxx.lanl.gov/abs/1403.1212}{{\tt
  1403.1212}}].

\bibitem{Profumo:2008ms}
S.~Profumo, {\it {Dissecting cosmic-ray electron-positron data with Occam's
  Razor: the role of known Pulsars}},  {\em Central Eur. J. Phys.} {\bf 10}
  (2011) 1--31, [\href{http://xxx.lanl.gov/abs/0812.4457}{{\tt 0812.4457}}].

\bibitem{Hooper:2008kg}
D.~Hooper, P.~Blasi, and P.~D. Serpico, {\it {Pulsars as the Sources of High
  Energy Cosmic Ray Positrons}},  {\em JCAP} {\bf 0901} (2009) 025,
  [\href{http://xxx.lanl.gov/abs/0810.1527}{{\tt 0810.1527}}].

\bibitem{Grasso:2009ma}
{\bf Fermi-LAT} Collaboration, D.~Grasso {\em et.~al.}, {\it {On possible
  interpretations of the high energy electron-positron spectrum measured by the
  Fermi Large Area Telescope}},  {\em Astropart. Phys.} {\bf 32} (2009)
  140--151, [\href{http://xxx.lanl.gov/abs/0905.0636}{{\tt 0905.0636}}].

\bibitem{Aguilar:2019owu}
{\bf AMS} Collaboration, M.~Aguilar {\em et.~al.}, {\it {Towards Understanding
  the Origin of Cosmic-Ray Positrons}},  {\em Phys. Rev. Lett.} {\bf 122}
  (2019), no.~4 041102.

\bibitem{Abeysekara:2017old}
{\bf HAWC} Collaboration, A.~U. Abeysekara {\em et.~al.}, {\it {Extended
  gamma-ray sources around pulsars constrain the origin of the positron flux at
  Earth}},  {\em Science} {\bf 358} (2017), no.~6365 911--914,
  [\href{http://xxx.lanl.gov/abs/1711.06223}{{\tt 1711.06223}}].

\bibitem{Farzan:2019qdm}
Y.~Farzan and M.~Rajaee, {\it {Dark Matter Decaying into Millicharged Particles
  as a Solution to AMS 02 Positron Excess}},  {\em JCAP} {\bf 1904} (2019),
  no.~04 040, [\href{http://xxx.lanl.gov/abs/1901.11273}{{\tt 1901.11273}}].

\bibitem{Campos:2017odj}
M.~D. Campos, F.~S. Queiroz, C.~E. Yaguna, and C.~Weniger, {\it {Search for
  right-handed neutrinos from dark matter annihilation with gamma-rays}},  {\em
  JCAP} {\bf 1707} (2017), no.~07 016,
  [\href{http://xxx.lanl.gov/abs/1702.06145}{{\tt 1702.06145}}].

\bibitem{Batell:2017rol}
B.~Batell, T.~Han, and B.~Shams Es~Haghi, {\it {Indirect Detection of Neutrino
  Portal Dark Matter}},  {\em Phys. Rev.} {\bf D97} (2018), no.~9 095020,
  [\href{http://xxx.lanl.gov/abs/1704.08708}{{\tt 1704.08708}}].

\bibitem{Gelmini:1982rr}
G.~B. Gelmini, S.~Nussinov, and M.~Roncadelli, {\it {Bounds and Prospects for
  the Majoron Model of Left-handed Neutrino Masses}},  {\em Nucl. Phys.} {\bf
  B209} (1982) 157--173.

\bibitem{Gelmini:1984pe}
G.~Gelmini, D.~N. Schramm, and J.~W.~F. Valle, {\it {Majorons: A Simultaneous
  Solution to the Large and Small Scale Dark Matter Problems}},  {\em Phys.
  Lett.} {\bf 146B} (1984) 311--317.

\bibitem{Santamaria:1986kg}
A.~Santamaria, J.~Bernabeu, and A.~Pich, {\it {Neutrino Masses, Majorons and
  Muon Decay}},  {\em Phys. Rev.} {\bf D36} (1987) 1408.

\bibitem{Choi:1991aa}
K.~Choi and A.~Santamaria, {\it {17-KeV neutrino in a singlet - triplet majoron
  model}},  {\em Phys. Lett.} {\bf B267} (1991) 504--508.

\bibitem{Berezinsky:1993fm}
V.~Berezinsky and J.~W.~F. Valle, {\it {The KeV majoron as a dark matter
  particle}},  {\em Phys. Lett.} {\bf B318} (1993) 360--366,
  [\href{http://xxx.lanl.gov/abs/hep-ph/9309214}{{\tt hep-ph/9309214}}].

\bibitem{Chang:2014lxa}
W.-F. Chang and J.~N. Ng, {\it {Minimal model of Majoronic dark radiation and
  dark matter}},  {\em Phys. Rev.} {\bf D90} (2014), no.~6 065034,
  [\href{http://xxx.lanl.gov/abs/1406.4601}{{\tt 1406.4601}}].

\bibitem{Queiroz:2014yna}
F.~S. Queiroz and K.~Sinha, {\it {The Poker Face of the Majoron Dark Matter
  Model: LUX to keV Line}},  {\em Phys. Lett.} {\bf B735} (2014) 69--74,
  [\href{http://xxx.lanl.gov/abs/1404.1400}{{\tt 1404.1400}}].

\bibitem{Boucenna:2014uma}
S.~M. Boucenna, S.~Morisi, Q.~Shafi, and J.~W.~F. Valle, {\it {Inflation and
  majoron dark matter in the seesaw mechanism}},  {\em Phys. Rev.} {\bf D90}
  (2014), no.~5 055023, [\href{http://xxx.lanl.gov/abs/1404.3198}{{\tt
  1404.3198}}].

\bibitem{Ma:2017xxj}
E.~Ma and M.~Maniatis, {\it {Pseudo-Majoron as Light Mediator of Singlet Scalar
  Dark Matter}},  {\em JHEP} {\bf 07} (2017) 140,
  [\href{http://xxx.lanl.gov/abs/1704.06675}{{\tt 1704.06675}}].

\bibitem{Garcia-Cely:2017oco}
C.~Garcia-Cely and J.~Heeck, {\it {Neutrino Lines from Majoron Dark Matter}},
  \href{http://xxx.lanl.gov/abs/1701.07209}{{\tt 1701.07209}}.

\bibitem{Brune:2018sab}
T.~Brune and H.~Päs, {\it {Massive Majorons and constraints on the
  Majoron-neutrino coupling}},  {\em Phys. Rev.} {\bf D99} (2019), no.~9
  096005, [\href{http://xxx.lanl.gov/abs/1808.08158}{{\tt 1808.08158}}].

\bibitem{Ando:2015qda}
S.~Ando and K.~Ishiwata, {\it {Constraints on decaying dark matter from the
  extragalactic gamma-ray background}},  {\em JCAP} {\bf 1505} (2015), no.~05
  024, [\href{http://xxx.lanl.gov/abs/1502.02007}{{\tt 1502.02007}}].

\bibitem{Massari:2015xea}
A.~Massari, E.~Izaguirre, R.~Essig, A.~Albert, E.~Bloom, and G.~A.
  Gómez-Vargas, {\it {Strong Optimized Conservative $Fermi$-LAT Constraints on
  Dark Matter Models from the Inclusive Photon Spectrum}},  {\em Phys. Rev.}
  {\bf D91} (2015), no.~8 083539,
  [\href{http://xxx.lanl.gov/abs/1503.07169}{{\tt 1503.07169}}].

\bibitem{Cohen:2016uyg}
T.~Cohen, K.~Murase, N.~L. Rodd, B.~R. Safdi, and Y.~Soreq, {\it {γ -ray
  Constraints on Decaying Dark Matter and Implications for IceCube}},  {\em
  Phys. Rev. Lett.} {\bf 119} (2017), no.~2 021102,
  [\href{http://xxx.lanl.gov/abs/1612.05638}{{\tt 1612.05638}}].

\bibitem{Genolini:2019ewc}
Y.~Génolini {\em et.~al.}, {\it {Cosmic-ray transport from AMS-02 boron to
  carbon ratio data: Benchmark models and interpretation}},  {\em Phys. Rev.}
  {\bf D99} (2019), no.~12 123028,
  [\href{http://xxx.lanl.gov/abs/1904.08917}{{\tt 1904.08917}}].

\bibitem{Ando:2016ang}
S.~Ando and K.~Ishiwata, {\it {Constraining particle dark matter using local
  galaxy distribution}},  {\em JCAP} {\bf 1606} (2016), no.~06 045,
  [\href{http://xxx.lanl.gov/abs/1604.02263}{{\tt 1604.02263}}].

\end{thebibliography}\endgroup

\end{document}